\documentclass[sort&compress]{elsarticle}
\usepackage{graphicx}
\usepackage{subfigure}
\usepackage{dcolumn}
\usepackage{bm}
\usepackage{color}
\begin{document}
\title{Quantum interference induced by initial system-environment correlations}

\author[rvt]{Zhong-Xiao Man}

\author[focal,els]{Andrea Smirne\corref{cor1}}
\ead{andrea.smirne@unimi.it}

\author[rvt]{Yun-Jie Xia}

\author[focal,els]{Bassano Vacchini}

\address[rvt]{Shandong Provincial Key Laboratory of Laser
Polarization and Information Technology, Department of Physics, Qufu
Normal University, Qufu 273165, China}

\address[focal]{Dipartimento di Fisica, Universit\`{a} degli Studi di Milano, Via Celoria 16, I-20133 Milan, Italy}

\address[els]{INFN, Sezione di Milano, Via Celoria 16, I-20133 Milan, Italy}

\cortext[cor1]{Corresponding author.}

\begin{abstract}
We investigate the quantum interference induced by a
relative phase in the correlated initial state of a system which consists in a
two-level atom interacting with a damped mode of the radiation
field. We show that the initial relative phase has significant  effects on both the evolution of the atomic excited-state population and
the information flow between the atom and the reservoir, as quantified by the trace distance.
Furthermore, by considering two two-level atoms interacting with a {}{common} damped mode of the radiation field,
we highlight how initial relative phases can affect the subsequent entanglement dynamics.
\end{abstract}


\maketitle

\section{Introduction}
Understanding of the time-evolution of open systems
is of great relevance not only in the foundations of quantum mechanics,
but also in the rapidly developing quantum technologies \cite{open, qci}.
In most cases, the initial state of the open system is assumed
uncorrelated from its environment, so that the evolution of the open system
can be described by a family of completely positive trace preserving (CPT) reduced dynamical maps.
Nevertheless, in many concrete experimental situations the investigated system
is unavoidably correlated with the environment also at the initial time,
especially in the case of systems which are strongly coupled to the reservoir.
Therefore, initial correlations between the system and the reservoir represent a fundamental issue, both
from a theoretical and an experimental point of view \cite{ini1,ini2,ini4,ini5,
ini7,ini10,ini11,ini12,ini13,ini14,ini15}.
In particular,
if there are initial correlations between an open system and the corresponding environment, the trace distance of two states
of the open system can grow over its initial value during the time-evolution \cite{trace1,trace3,JCM}, indicating that
the open system can access some information which is initially outside it.
Recently, the trace-distance growth of open-system states above its initial value has been experimentally
verified {by means of all-optical apparata, in which
the system under investigation consists of single photons emitted by quantum dots
or couples of entangled photons generated by parametric down conversion} \cite{trace2}.

In this paper, we focus on a general effect induced by initial system-en\-vi\-ron\-ment correlations in the
subsequent open-system dynamics. Namely, we show that the quantum interference
due to a relative phase in the correlated initial total state
plays an important role in several
aspects of the dynamics of
the open system. In particular, we investigate a system composed of a two-level atom
interacting with a mode of the radiation field, which is in turn
coupled to a damping reservoir.
First, we study the dynamics of atomic excited-state
population, thus showing how the quantum interference
can modify in different ways the system's energy gain
from the reservoir.
Then, we take into account the evolution of the trace distance between atomic states,
characterizing the information flow between the
atom and the reservoir. It is shown that the initial relative phase plays a basic role in order to maximize the
increase of the trace distance above its initial value, i.e., the amount of information that
flows from the environment
to the open system in the course of the dynamics. Finally, we generalize our model,
by taking into account two two-level atoms interacting with a {}{common} damped mode of the radiation field.
We study how relative phases in a correlated initial atom-mode state influence the dynamics of atomic entanglement,
and in particular the steady-state entanglement.

\section{Model and solution}\label{sec:ms}
In this paper, we consider the damped Jaynes-Cummings model \cite{open},
namely a two level atom interacting via the Jaynes-Cummings Hamiltonian with
a mode of the radiation field, that is in turn coupled to a damping reservoir.
Our aim is to investigate how
initial correlations between the atom and the mode influence the subsequent dynamics of the atom and,
in particular, we focus on the role of the relative phase in the correlated initial state. Thus, we
are not assuming an initial vacuum state of the mode.
In the following, we will use the Lindblad master equation for the atom-mode system given by \cite{cohen}
\begin{equation}\label{mast}
\frac{d\rho(t)}{dt}=-i[H,\rho(t)]-\frac{\Gamma}{2}[b^{+}b\rho(t)-2b\rho(t) b^{+}+\rho(t) b^{+}b],
\end{equation}
with
$
H=\omega _{0}{\sigma}_{+}{\sigma}_{-}
+\omega _{c}{b}^{+}{b}
+\Omega({\sigma}_{-}{b}^{+}+
{\sigma}_{+}{b}),
$
where $\rho$ represents the density matrix of the system formed by the atom and the
mode, ${\sigma}_{\pm}$ are the raising and lowering operators and $\omega _{0}$ is the transition
frequency of the atom. Moreover, $b$ ($b^{\dagger}$) is the
annihilation (creation) operator of the mode, with frequency $\omega_c$, $\Omega$
is the coupling constant between the atom and the mode, and, finally, $\Gamma$
is the damping rate of the mode due to its interaction with the dissipative reservoir.
We focus on the resonant case, i.e. $\omega_0 = \omega_c \equiv \omega$. Note that this
master equation, often introduced on the basis of a phenomenological approach, can be microscopically justified for a zero temperature flat reservoir \cite{damp}
{relying on the Born-Markov approximation \cite{open}. It then provides
a description of the atom-mode dynamics on a coarse grained time scale, which does not resolve the decay time of correlation
functions of the damping reservoir.}
Finally, let us introduce the basis $|\alpha\rangle_{A} \otimes |n\rangle_{M} \equiv |\alpha, n\rangle_{AM}$, with $\alpha = e,g$
labeling the states of the two-level atom, and $n = 0,1,2, \ldots$ the number states of
the field mode.

In the following, we first take into account a correlated initial atom-mode pure state $
\rho_{AM}^{(1)}(0)=\left|\Psi(0)\right\rangle_{AM}\left\langle\Psi(0)\right|$, with
\begin{equation}\label{corr}
\left|\Psi(0)\right\rangle_{AM}=C_{1}(0)\left|e,0\right\rangle_{AM}+
C_{2}(0)\left|g,1\right\rangle_{AM}.
\end{equation}
The reduced states of the atom and the mode are, respectively,
$\rho_{A}^{(1)}(0)=|C_{1}(0)|^2\left|e\right\rangle\left\langle e\right|
+|C_{2}(0)|^2\left|g\right\rangle\left\langle g\right|$ and
$\rho_{M}^{(1)}(0)=|C_{1}(0)|^2\left|0\right\rangle\left\langle 0\right|
+|C_{2}(0)|^2\left|1\right\rangle\left\langle 1\right|$.
To make a comparison with the uncorrelated situation, we then consider a product
initial atom-mode state in the form
\begin{equation}\label{prod}
\rho_{AM}^{(2)}(0)=\rho_{A}^{(2)}(0)\otimes\rho_{M}^{(2)}(0),
\end{equation}
with
$\rho_{A}^{(2)}(0)=|B_{1}(0)|^2\left|e\right\rangle\left\langle e\right|
+|B_{2}(0)|^2\left|g\right\rangle\left\langle g\right|$
and $\rho_{M}^{(2)}(0)=\rho_{M}^{(1)}(0)$. Note that the initial atom-mode states
$\rho_{AM}^{(1)}(0)$ and $\rho_{AM}^{(2)}(0)$ have the same
reduced states for the mode, but they may have different reduced states
for the atom.

Now, consider the correlated initial state $\rho_{AM}^{(1)}(0)$. Since there is only one
excitation in the total system, we can make
the ansatz that the atom-mode state at time $t$ is of the form \cite{pseu1, pseudo}
\begin{equation}\label{eq:ansatz}
\rho^{(1)}_{AM}(t) = (1-\lambda(t)) |\psi(t)\rangle_{AM} \langle \psi(t)| + \lambda(t) |g, 0\rangle_{AM}\langle g,0|,
\end{equation}
with $0\leq \lambda(t) \leq 1$, $\lambda(0)=0$, and
$
|\psi(t)\rangle_{AM}= C_1(t) |e, 0\rangle_{AM} + C_2(t) |g, 1 \rangle_{AM}.
$
In the spirit of \cite{pseu1}, it is convenient to introduce the unnormalized state vector
\begin{equation}\label{eq:unno}
|\widetilde{\psi}(t)\rangle_{AM} \equiv {\sqrt{(1-\lambda(t))}}|\psi(t)\rangle_{AM} =  \widetilde{C}_1(t) |e, 0\rangle_{AM} + \widetilde{C}_2(t) |g, 1 \rangle_{AM},
\end{equation}
where $\widetilde{C}_k(t) \equiv {\sqrt{(1-\lambda(t))}}C_k(t)$, for $k=1,2$, and therefore the atom-mode state at time $t$ can be written as
$\rho^{(1)}_{AM}(t) = |\widetilde{\psi}(t)\rangle_{AM} \langle \widetilde{\psi}(t)| + \lambda(t) |g, 0\rangle_{AM}\langle g,0|$.
Due to Eq.(\ref{mast}) the dynamics of the unnormalized atom-mode
state in Eq.~(\ref{eq:unno}) is determined by \cite{pseu1}:
\begin{eqnarray}
i\frac{d}{dt}\widetilde{C}_{1}(t)&=&\omega \widetilde{C}_{1}(t)+\Omega \widetilde{C}_{2}(t),\nonumber\\
i\frac{d}{dt}\widetilde{C}_{2}(t)&=&\left(\omega- i \frac{\Gamma}{2}\right)\widetilde{C}_{2}(t)+\Omega \widetilde{C}_{1}(t),\label{eq:garra}
\end{eqnarray}
while $\lambda(t)$ satisfies
$d \lambda(t) / d t = \Gamma | \widetilde{C}_2(t) |^2$.
By means of the Laplace transformation, together with the initial condition
$\left|\Psi(0)\right\rangle_{AM}$ given by Eq.~(\ref{corr}),
we get the analytical expression $\widetilde{C}_{1}(t) =\mu(t)C_{1}(0)-i\nu(t)C_{2}(0)$,
with
\begin{eqnarray}
\mu(t)&=&e^{-(\frac{\Gamma}{4}+ i \omega) t}\left[\cosh(\frac{a t}{2})+\frac{\Gamma}{2 a}
\sinh(\frac{a t}{2})\right],\label{mu} \\
\nu(t)&=&e^{-(\frac{\Gamma}{4}+ i \omega) t}\left[\frac{2\Omega}{a}
\sinh(\frac{a t}{2})\right],\label{nu}
\end{eqnarray}
and $a=\sqrt{(\Gamma/2)^{2}-4\Omega^{2}}$.
One may distinguish two different dynamical regimes via
$\Gamma$ and $\Omega$; namely, we will identify the case
{$\Gamma> 2\Omega$ as the weak coupling regime and
$\Gamma< 2\Omega$ as the strong coupling regime} \cite{pseudo}. From Eqs.~(\ref{eq:ansatz}) and (\ref{eq:unno}),
one has that the reduced state of the atom at time $t>0$
can be expressed as
\begin{equation}\label{At1}
\rho_{A}^{(1)}(t)=|C_{e}^{(1)}(t)|^{2}\left|e\right\rangle\left\langle e\right|
+(1-|C_{e}^{(1)}(t)|^{2})\left|g\right\rangle\left\langle g\right|,
\end{equation}
with
$|C_{e}^{(1)}(t)|^{2}\equiv|\widetilde{C}_{1}(t)|^{2}$.

Let us now take into account the uncorrelated initial atom-mode state $\rho_{AM}^{(2)}(0)$. It can be written as the
mixture of four pure states, i.e., $\left|e,0\right\rangle\left\langle e,0\right|$,
$\left|g,1\right\rangle\left\langle g,1\right|$,
$\left|e,1\right\rangle\left\langle e,1\right|$ and
$\left|g,0\right\rangle\left\langle g,0\right|$.
For later convenience, we separately evaluate
the different contributions given by these four terms to
the probability of
the atom being in excited state at a time $t$. By the previous analysis, it is clear that the contributions
due to
$\left|e,0\right\rangle\left\langle e,0\right|$
and $\left|g,1\right\rangle\left\langle g,1\right|$ are just, respectively, {$|\mu(t)|^{2}$
and $|\nu(t)|^{2}$}, see Eqs.(\ref{mu}) and (\ref{nu}).
The term $\left|g,0\right\rangle\left\langle g,0\right|$ is invariant, while for the contribution of $\left|e,1\right\rangle\left\langle e,1\right|$
we have to come back to the master equation (\ref{mast}) and set the initial condition $\rho(0) =|e, 1\rangle\langle e,1|$, thus
getting the system of equations:
\begin{eqnarray}
\dot{\varrho}_{11}(t) &=& -i[\sqrt{2}\varrho_{21}(t)\Omega-\sqrt{2}\varrho_{12}(t)\Omega]-\Gamma\varrho_{11}(t)\nonumber \\
\dot{\varrho}_{12}(t) &=& -i[\varrho_{12}(t)\omega_{0}-\varrho_{12}(t)\omega_{c}+\sqrt{2}\varrho_{22}(t)\Omega
-\sqrt{2}\varrho_{11}(t)\Omega]-\frac{3\Gamma}{2}\varrho_{12}(t) \nonumber\\
\dot{\varrho}_{22}(t) &=& -i[\sqrt{2}\varrho_{12}(t)\Omega-\sqrt{2}\varrho_{21}(t)\Omega]-2\Gamma \varrho_{22}(t) \nonumber\\
\dot{\varrho}_{33}(t) &=& -i[\varrho_{43}(t)\Omega-\varrho_{34}(t)\Omega]+\Gamma \varrho_{11}(t) \nonumber\\
\dot{\varrho}_{34}(t) &=& -i[\varrho_{34}(t)\omega_{0}+\varrho_{44}(t)\Omega
-\varrho_{34}(t)\omega_{c}-\varrho_{33}(t)\Omega]-\frac{\Gamma}{2}[\varrho_{34}(t)-2\sqrt{2}\varrho_{12}(t)] \nonumber\\
\dot{\varrho}_{44}(t) &=& -i[\varrho_{44}(t)\omega_{c}+\varrho_{34}(t)\Omega-\varrho_{43}(t)\Omega]
-\frac{\Gamma}{2}[2\varrho_{44}(t)-4\varrho_{22}(t)].
\end{eqnarray}
We indicated the atom-mode state at time $t$ as $\varrho(t)$ to emphasize that it
corresponds to the specific above-mentioned initial condition,
and the matrix elements of $\varrho(t)$ are expressed with respect to
the atom-mode states
$\{\left|\widetilde{0}\right\rangle=\left|g,0\right\rangle,
\left|\widetilde{1}\right\rangle=\left|e,1\right\rangle,
\left|\widetilde{2}\right\rangle=\left|g,2\right\rangle,
\left|\widetilde{3}\right\rangle=\left|e,0\right\rangle,
\left|\widetilde{4}\right\rangle=\left|g,1\right\rangle\}$,
with the notation $\varrho_{kl}(t)\equiv \left\langle \widetilde{k}\right|\varrho(t)\left|
\widetilde{l}\right\rangle$.
By solving these equations, we get
the probability $\varrho_{11}(t)+\varrho_{33}(t)$ for atomic excited-state occupation
induced by the term $\left|e,1\right\rangle\left\langle e,1\right|$.
Summarizing, the reduced state of the atom at time $t>0$ for the initial atom-mode state $\rho_{AM}^{(2)}(0)$
reads
$\rho_{A}^{(2)}(t)=|C_{e}^{(2)}(t)|^2\left|e\right\rangle\left\langle e\right|
+(1-|C_{e}^{(2)}(t)|^2)\left|g\right\rangle\left\langle g\right|
$
with
\begin{equation}\label{At3}
 |C_{e}^{(2)}(t)|^2=
|B_{1}(0)C_{1}(0)\mu(t)|^{2}
+|B_{2}(0)C_{2}(0)\nu(t)|^{2}
+|B_{1}(0)C_{2}(0)|^{2}[\varrho_{1 1}(t)+\varrho_{3 3}(t)].
\end{equation}

\section{Dynamics of atomic excited-state population}\label{sec:doa}
In this work, we are concerned with the specific effects of initial atom-mode correlations
on atomic dynamics {and, in particular,
we stress the role of the relative phase in the correlated initial atom-mode state. Thus, let us} reexpress $C_{1}(0)$ and $C_{2}(0)$ in Eq. (\ref{corr}) as
$
C_{1}(0)=\mathcal{C}_{1}(0)$ and $C_{2}(0)=\mathcal{C}_{2}(0)e^{i\theta},
$
where $\mathcal{C}_{1}(0)$ and $\mathcal{C}_{2}(0)$ are real numbers and $\theta\in[0,2\pi]$.
For the correlated initial state, the atomic
excited-state population is given by, see Eq.~(\ref{At1}),
\begin{equation}\label{exc-pop}
|C_{e}^{(1)}(t)|^{2}=|\mu(t)|^2\mathcal{C}_{1}^{2}(0)+|\nu(t)|^2\mathcal{C}_{2}^{2}(0)
+2 \mu(t)\nu^*(t)\mathcal{C}_{1}(0)\mathcal{C}_{2}(0)\sin\theta,
\end{equation}
with $\mu(t)$ and $\nu(t)$ as in Eqs. (\ref{mu}) and (\ref{nu}), respectively.
The first term represents the
transfer of excitation that is initially in the atom with probability $\mathcal{C}_{1}^{2}(0)$,
while the second term represents the transfer of excitation that is initially in the mode with probability
$\mathcal{C}_{2}^{2}(0)$. These two processes coexist in the whole course of evolution
and, indeed, quantum interference can be induced between them,
which is just denoted by the third term in Eq. (\ref{exc-pop}).
Obviously, the constructive (destructive)
quantum interference corresponds to $\theta\in(0,\pi)$ ($\theta\in(\pi,2\pi)$), {while for $\theta=0,\pi,2\pi$
there is no quantum interference}.
For convenience, we define a rescaled probability of the atomic excited-state population as
$
\tilde{P}_{e}(t)=|C_{e}^{(k)}(t)/C_{e}^{(k)}(0)|^{2},
$
where $k=1,2$ refer to, respectively, the initially correlated and uncorrelated states of
the total system.

\begin{figure}[t]
\centering
\begin{minipage}[b]{2.5in}
\centering
\includegraphics[width=2.5in]{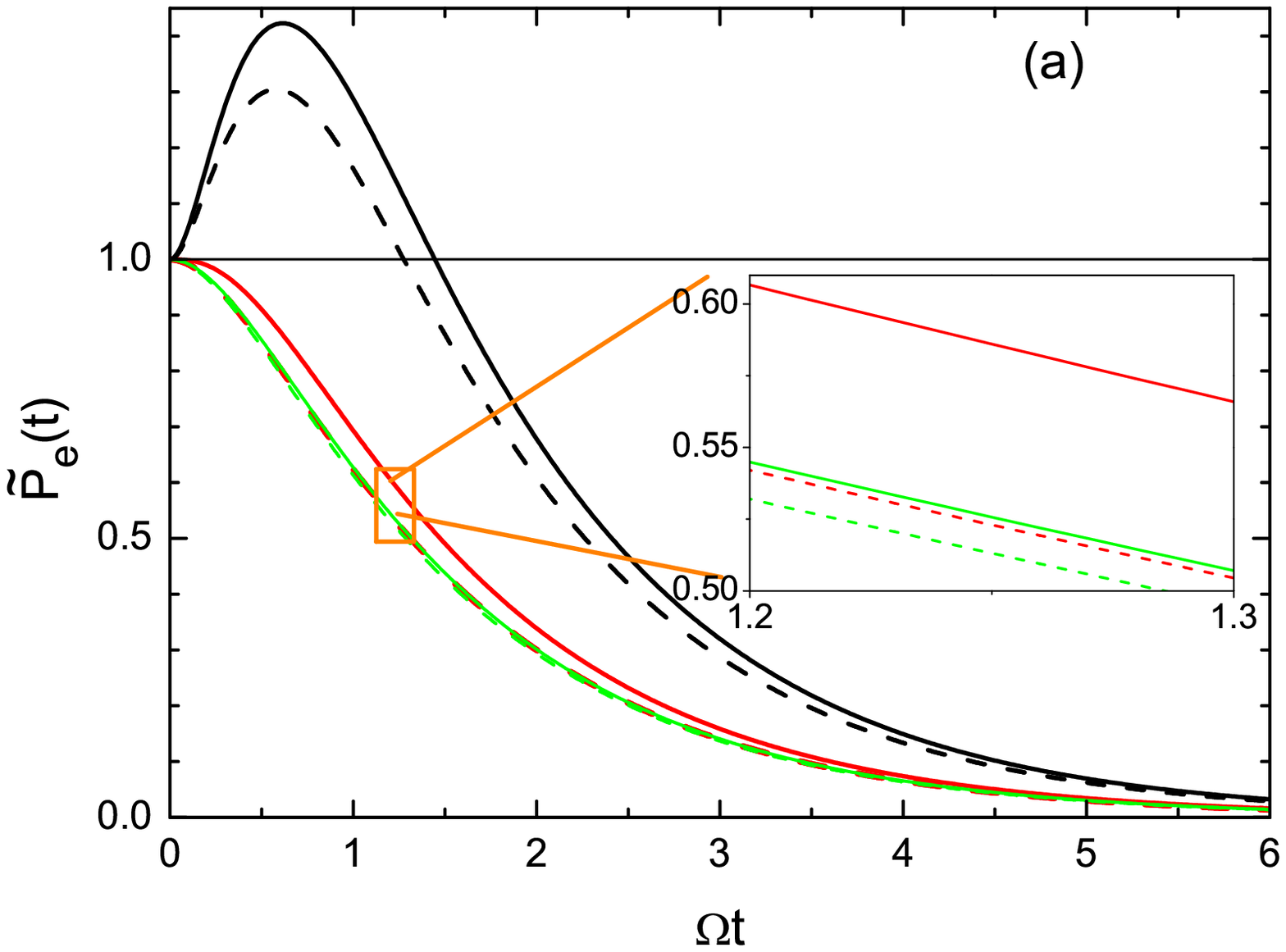}
\end{minipage}%
\begin{minipage}[b]{2.5in}
\centering
\includegraphics[width=2.5in]{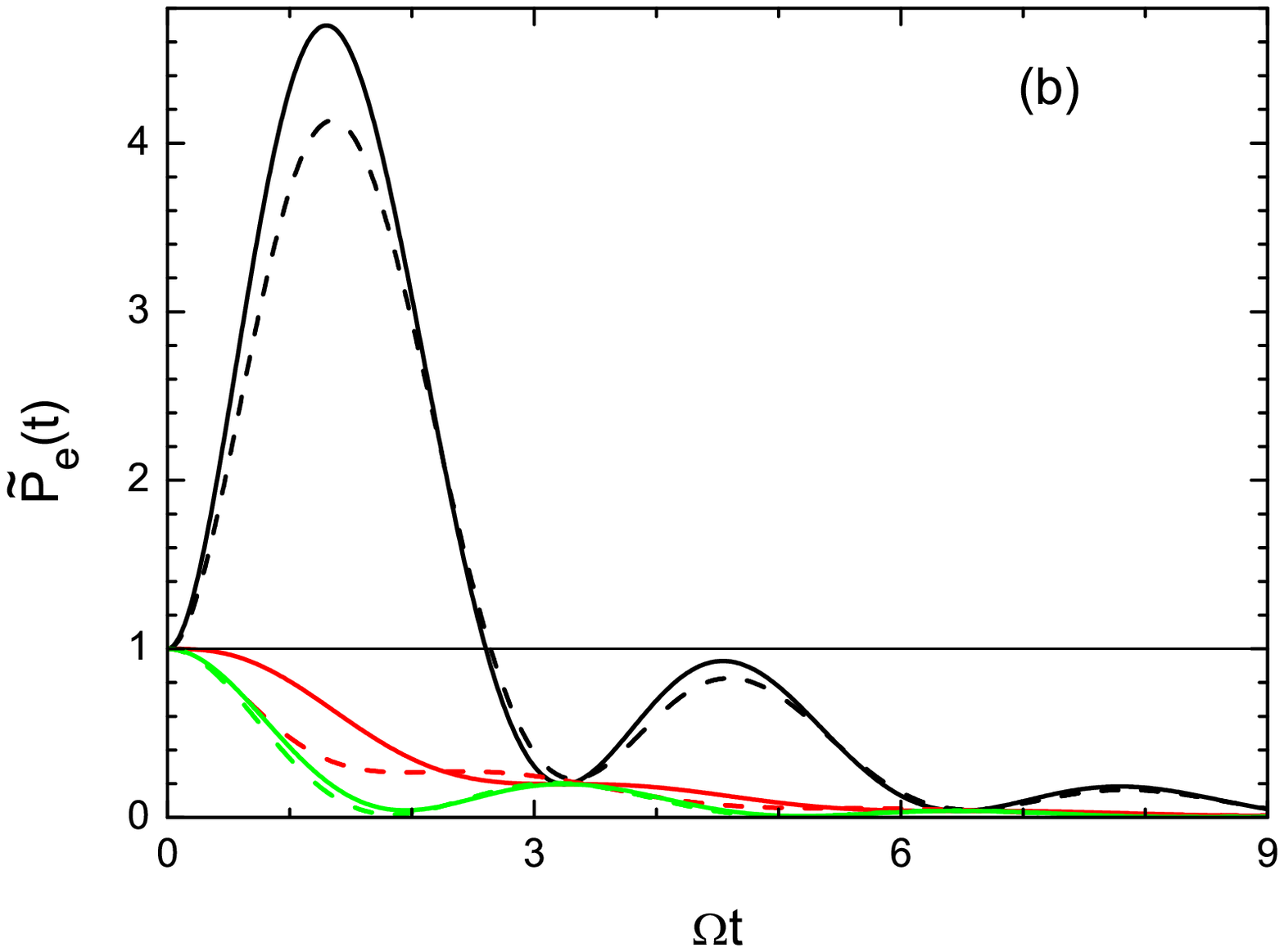}
\end{minipage}\\
\centering
\begin{minipage}[b]{2.5in}
\centering
\includegraphics[width=2.5in]{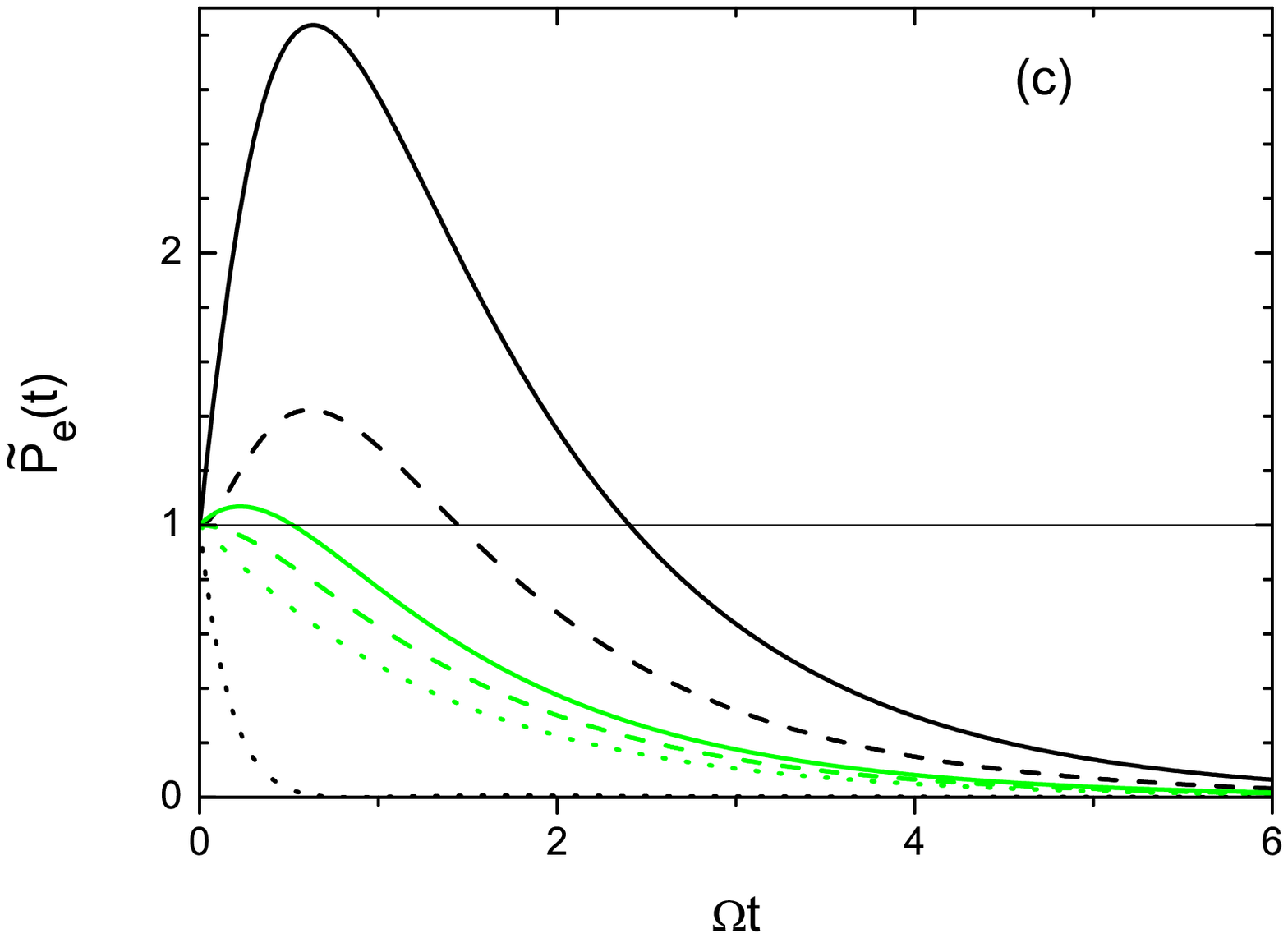}
\end{minipage}%
\begin{minipage}[b]{2.5in}
\centering
\includegraphics[width=2.5in]{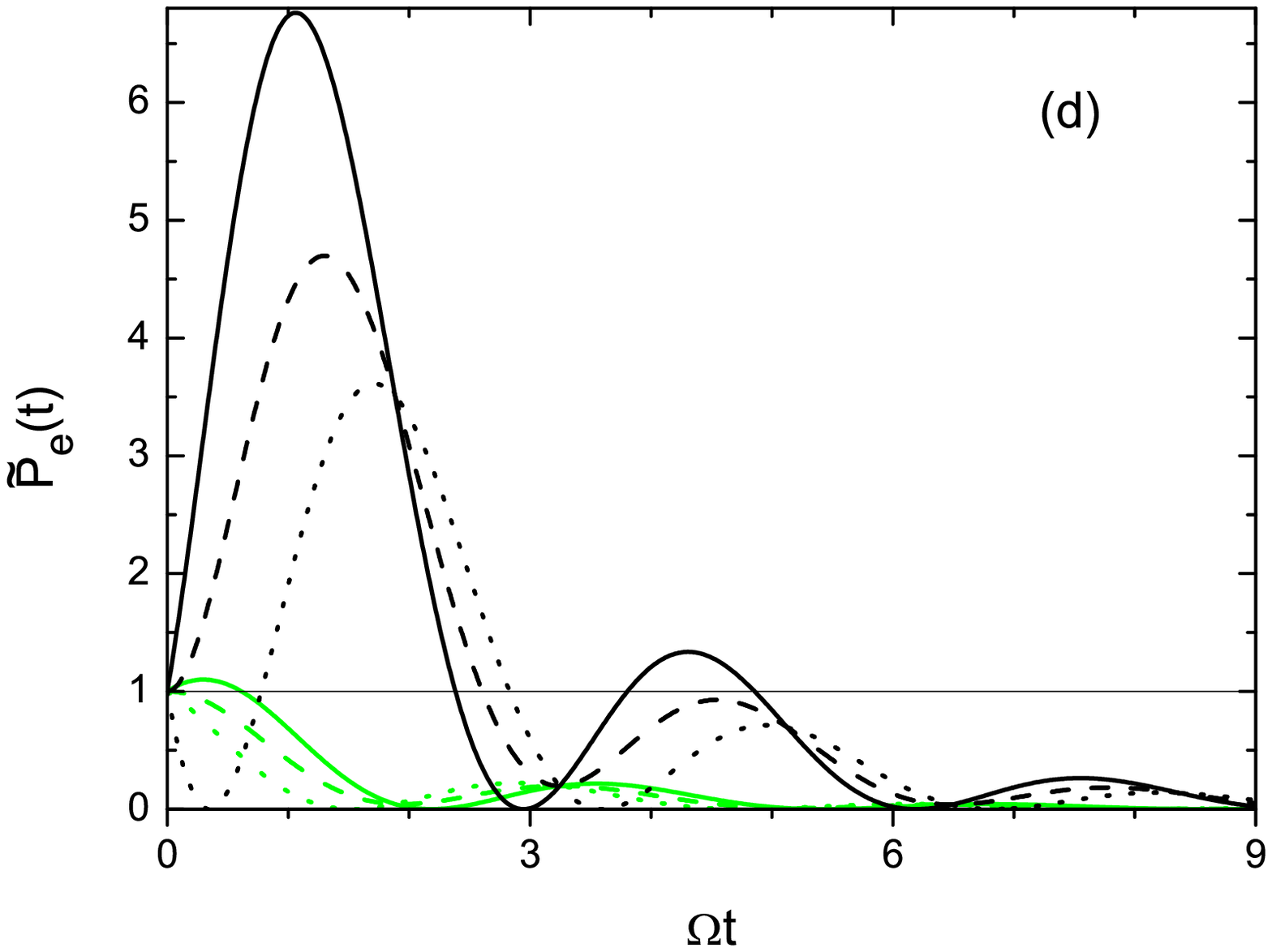}
\end{minipage}
\caption{(a,b): Time-evolution of $\tilde{P}_{e}(t)$ 
for the correlated initial state $\rho_{AM}^{(1)}(0)$ with $\theta=0$ (solid lines)
and for the uncorrelated initial state $\rho_{AM}^{(2)}(0)$ (dashed lines)
in (a) weak ($\Gamma/\Omega=6$) and (b) strong ($\Gamma/\Omega=1$) coupling regimes.
(c,d): Time-evolution of $\tilde{P}_{e}(t)$
for $\rho_{AM}^{(1)}(0)$ under different relative phases in terms of $\theta=\pi/2$ (solid lines),
$\theta=0$ (dashed lines) and $\theta=3\pi/2$ (dotted lines)
in (c) weak ($\Gamma/\Omega=6$) and (d) strong ($\Gamma/\Omega=1$) coupling regimes.
The other parameters are $\mathcal{C}_{1}(0)=\sqrt{\frac{1}{10}}<
\mathcal{C}_{2}(0)=\sqrt{\frac{9}{10}}$ (black line);
$\mathcal{C}_{1}(0)=\mathcal{C}_{2}(0)=\sqrt{\frac{1}{2}}$ (red line);
$\mathcal{C}_{1}(0)=\sqrt{\frac{9}{10}}>\mathcal{C}_{2}(0)=\sqrt{\frac{1}{10}}$ (green line),
and $B_{1}(0)=\mathcal{C}_{1}(0)$, $B_{2}(0)=\mathcal{C}_{2}(0)$.}\label{fig:1}
\end{figure}

First, we consider the {case without quantum interference}.
The uncorrelated initial state $\rho_{AM}^{(2)}(0)$ in Eq.~(\ref{prod}) is chosen as the tensor product of the marginals of the
correlated state $\rho_{AM}^{(1)}(0)$, so that the probability of atomic excited-state occupation is obtained by putting $B_{1}(0)=\mathcal{C}_{1}(0)$ and $B_{2}(0)=\mathcal{C}_{2}(0)$ into Eq.(\ref{At3}). 
{In Fig.\ref{fig:1} (a) and (b), we plot the time-evolution of
the rescaled probability of the atomic excited-state population
$\tilde{P}_{e}(t)$ for the initial states $\rho_{AM}^{(1)}(0)$ (solid lines) and $\rho_{AM}^{(2)}(0)$ (dashed lines)
and for different values of $\mathcal{C}_{1}(0)$ and $\mathcal{C}_{2}(0$)}.
In both weak and strong coupling regimes and for both initially correlated and
uncorrelated situations, $\tilde{P}_{e}(t)>1$ occurs when
$\mathcal{C}_{2}(0)>\mathcal{C}_{1}(0)$. We observe that the amplitude of the increase of
$\tilde{P}_{e}(t)$ from the initial value one in the presence of initial correlations is always larger than that without initial
correlations. Moreover, for $\mathcal{C}_{2}(0)\leq
\mathcal{C}_{1}(0)$ one can see that $\tilde{P}_{e}(t)$ in the presence of initial correlations
is still larger than without initial correlations (the inset in Fig.\ref{fig:1}
(a) shows an enlarged figure).
Initial correlations in the absence of interference can be
identified as a mechanism that increases the excitation transfer from
the mode to the atom and inhibit the opposite process.

Next, we explore the role of quantum interference in influencing
atomic excited-state population.
In Fig.\ref{fig:1} (c) and (d), we plot the time-evolution of $\tilde{P}_{e}(t)$
for the correlated initial state $\rho_{AM}^{(1)}(0)$  for different values of $\mathcal{C}_{1}(0)$ and $\mathcal{C}_{2}(0)$ and
for $\theta=\pi/2$  (solid lines), $\theta=3\pi/2$ (dotted lines) and $\theta=0$  (dashed lines), i.e. for, respectively, constructive, destructive  and no quantum interference.
{Indeed, for the corresponding product state $\rho_{AM}^{(2)}(0)$
the time-evolution of $\tilde{P}_{e}(t)$ is the same in all the different situations.
We observe that, in the case of constructive interference, the rescaled probability $\tilde{P}_{e}(t)$ for the correlated initial state 
always reaches values greater than one}, irrespective of
the ratio $\mathcal{C}_{2}(0)/\mathcal{C}_{1}(0)$.
Moreover, also for $\mathcal{C}_{2}(0)>\mathcal{C}_{1}(0)$ the amplitude growth of $\tilde{P}_{e}(t)$ has
largely increased as compared to the case $\theta=0$: {the constructive interference
boosts the energy flow from the mode to the atom and inhibits the opposite process}.
On the other hand, for a destructive interference,
$\tilde{P}_{e}(t)$ is always lesser than one in the weak coupling regime, irrespective of
the ratio $\mathcal{C}_{2}(0)/\mathcal{C}_{1}(0)$ {and, besides, the greater this ratio is,
the faster the atom decays. 
In addition,
in the strong coupling regime $\tilde{P}_{e}(t)$ decays on short time scales
and then, even if it
can increase up to values greater than one for
$\mathcal{C}_{2}(0)>\mathcal{C}_{1}(0)$ in the later evolution, its
maximum value is clearly smaller than that of $\theta=0$ and $\theta=\pi/2$.} 
Therefore, we can conclude that the destructive interference
boosts the energy flow from the atom to the mode.
Summarizing, as far as the ability to facilitate the energy flow from the mode to the
atom and restrain the opposite process is taken into account, initial correlations
turn out to be crucial because of interference effects.

\section{Dynamics of the trace distance between atomic states}\label{sec:dotd}
In this section, we investigate the influence of initial correlations and
relative phase on the dynamics of the trace
distance of atomic states.
Trace distance is one of the most employed quantifiers for the
distinguishability of quantum states. A change of the trace distance between
reduced-system states in the course of the dynamics indicates an information flow
between the open system and the
environment \cite{blp}. The trace distance $D(\rho_{1},\rho_{2})$
between two quantum
states $\rho_{1}$ and $\rho_{2}$ is defined as \cite{qci}
$
D(\rho_{1},\rho_{2})=\frac{1}{2}\|\rho_{1}-\rho_{2}\|_{1},
$
where $\|X\|_{1}=\mathrm{Tr}\sqrt{X^{\dag}X}$ is the trace norm of the operator $X$.
Any positive and
trace-preserving map $\mathcal{E}$ defined on the whole space of trace class
operators is a contraction for the trace distance, i.e.,
$
D(\mathcal{E}(\rho_{1}),\mathcal{E}(\rho_{2}))\leq D(\rho_{1},\rho_{2})$.
In the presence of initial correlations the contractivity may fail since
the trace distance $D(\rho^1_S(t), \rho^2_S(t))$ between two states of the open system,
$\rho^1_S(t)$ and $\rho^2_S(t)$ evolved from the initial total states $\rho^1_{SE}(0)$ and $\rho^2_{SE}(0)$,
can exceed its initial value in the course of time-evolution  \cite{trace1}.
One can further determine
an upper bound
to the growth of trace distance, which
is
$I(\rho_{SE}^{(1)}(0),\rho_{SE}^{(2)}(0))\equiv D(\rho_{SE}^{(1)}(0),\rho_{SE}^{(2)}(0))
-D(\rho_{S}^{1}(0),\rho_{S}^{2}(0))$. This quantity can be interpreted as
the relative information about the total initial states $\rho_{SE}^{(1)}(0)$
and $\rho_{SE}^{(2)}(0)$ that is initially
outside the open system, i.e., that is inaccessible for
local measurements performed on the open system only \cite{trace1}.

For the case at hand, the evolution of the total system composed by
the atom and the mode is not unitary, but it is given by the family of CPT maps
forming the semigroup fixed by Eq.~(\ref{mast}). Then, also in this case, from the contractivity of the trace
distance, one can immediately see that the increase
of the trace distance between atomic states over its initial value is bounded from above by
\begin{equation}
D(\rho_{A}^{(1)}(t),\rho_{A}^{(2)}(t))-D(\rho_{A}^{(1)}(0),\rho_{A}^{(2)}(0))\leq
I(\rho_{AM}^{(1)}(0),\rho_{AM}^{(2)}(0)).
\end{equation}
For the initial atom-mode states
$\rho_{AM}^{(1)}(0)$ given by Eq.~(\ref{corr}) and $\rho_{AM}^{(2)}(0)$ as in Eq.~(\ref{prod}),
the quantity $I(\rho_{AM}^{(1)}(0),\rho_{AM}^{(2)}(0))$
is given by
\begin{eqnarray}\label{bound}
&&I(\rho_{AM}^{(1)}(0),\rho_{AM}^{(2)}(0))=\frac{1}{2}(\sqrt{(|B_{2}(0)C_{1}(0)|^2-|B_{1}(0)C_{2}(0)|^2)^2+4|C_{1}(0)C_{2}(0)|^{2}}
\nonumber\\&&+|B_{2}(0)C_{1}(0)|^2+|B_{1}(0)C_{2}(0)|^2)
-||C_{2}(0)|^{2}-|B_{2}(0)|^{2}|.
\end{eqnarray}
When $I(\rho_{AM}^{(1)}(0),\rho_{AM}^{(2)}(0))>0$ the trace distance of atomic
states can increase above its initial value in the time-evolution, indicating
that the information initially outside the open system is flowing back to it
during the dynamics.

The evolution of the trace distance of open system's states in the presence of
initial correlations was investigated in Ref. \cite{JCM} for
a Jaynes-Cummings model without dissipation.
Here, we extend the study to the dissipative model, in order to examine the influence of both
the dissipation and the relative phase on the dynamics of the trace distance.
For the initial atom-mode states
$\rho_{AM}^{(1)}(0)$ in Eq.~(\ref{corr}) and $\rho_{AM}^{(2)}(0)$ in Eq.~(\ref{prod}), the trace distance between the corresponding atomic states at time
$t>0$ is simply given by
\begin{equation}\label{distance}
D(\rho_{A}^{(1)}(t),\rho_{A}^{(2)}(t))=||C_{e}^{(1)}(t)|^2-|C_{e}^{(2)}(t)|^2|.
\end{equation}

\begin{figure}[t]
\centering
\begin{minipage}[b]{2.5in}
\centering
\includegraphics[width=2.5in]{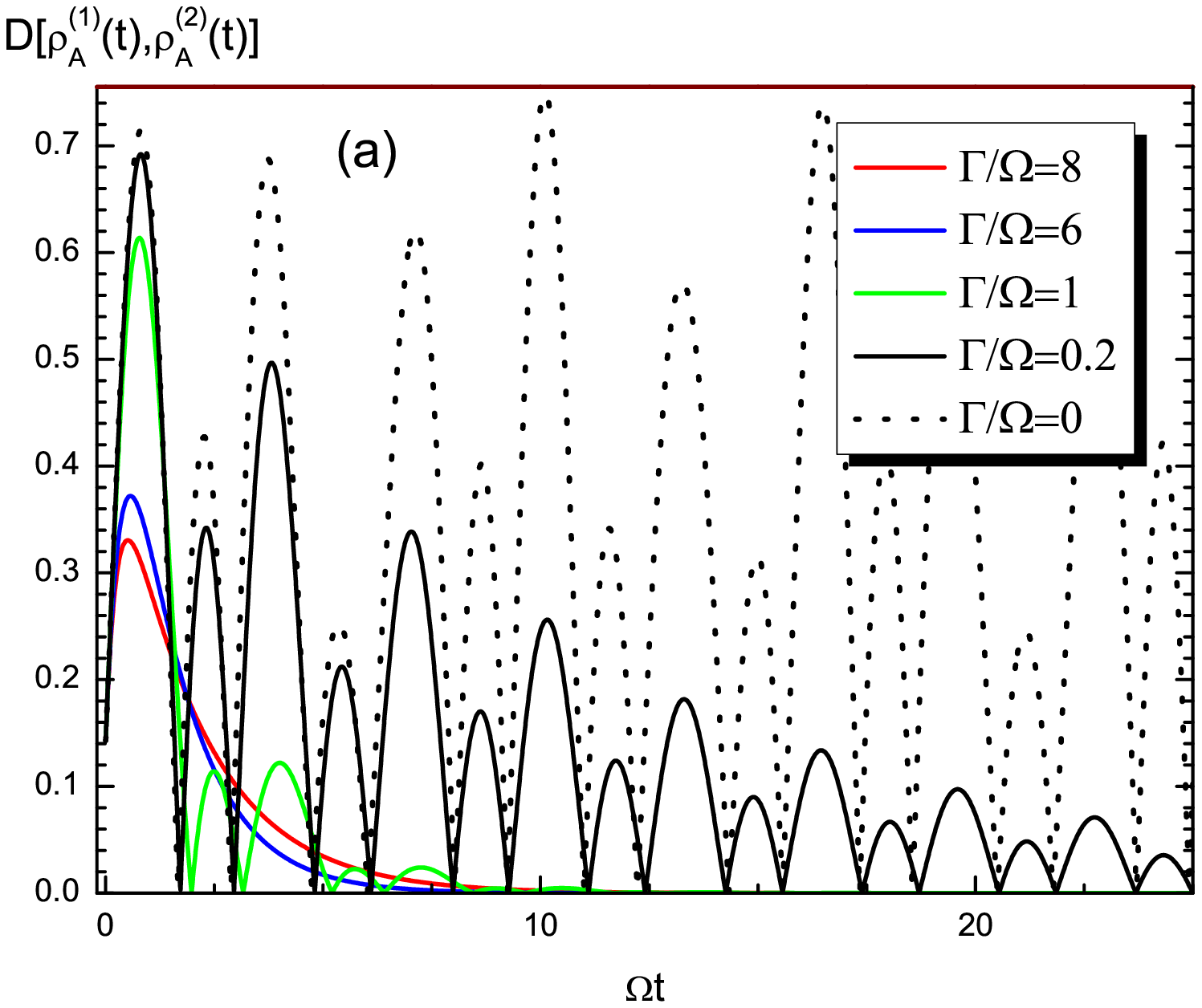}
\end{minipage}%
\begin{minipage}[b]{2.5in}
\centering
\includegraphics[width=2.5in]{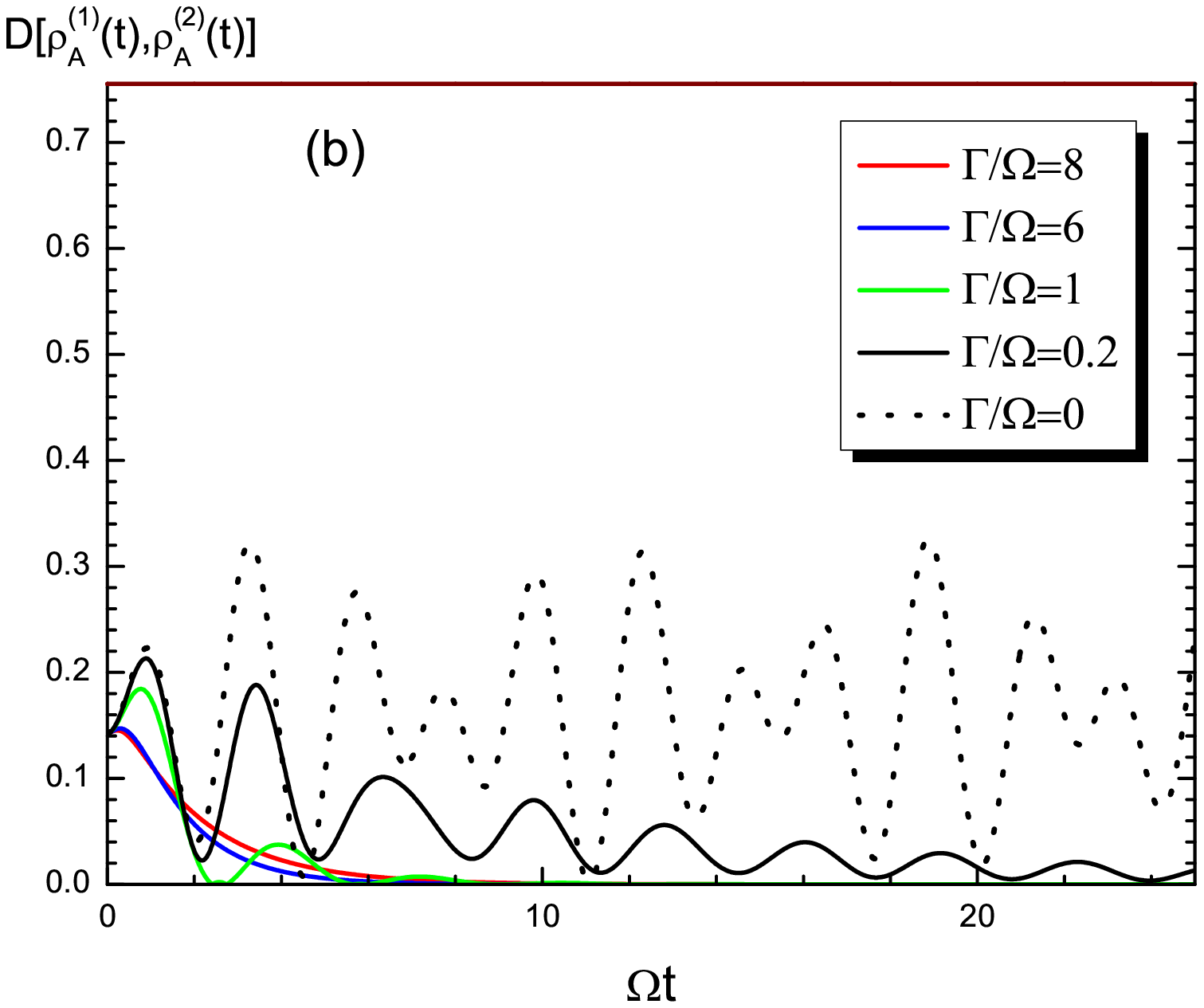}
\end{minipage}\\
\centering
\begin{minipage}[b]{2.5in}
\centering
\includegraphics[width=2.5in]{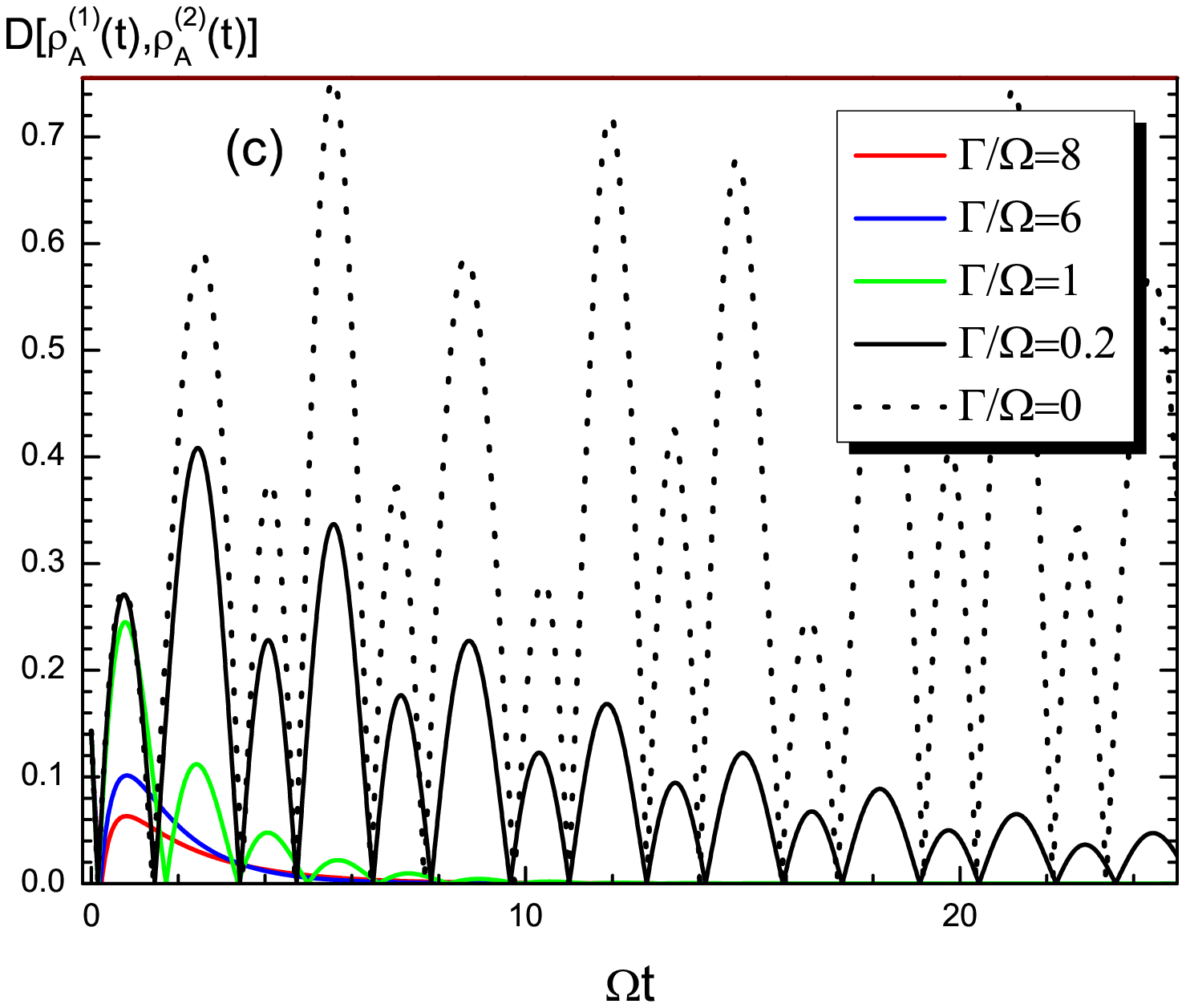}
\end{minipage}%
%\begin{figure}
%\hspace{-1.5cm}
%\begin{minipage}[b]{6.5in}
%\includegraphics[width=2in]{Fig2a}
%\includegraphics[width=2in]{Fig2b}
%\includegraphics[width=2in]{Fig2c}
%\end{minipage}
\caption{Plot of the trace distance $D(\rho_{A}^{(1)}(t),\rho_{A}^{(2)}(t))$
as a function of rescaled time $\Omega t$. In (a)
the parameters characterizing the correlated initial atom-mode state are
$C_{1}(0)=\sqrt{\frac{4}{7}}$, $C_{2}(0)=i\sqrt{\frac{3}{7}}$,
in (b) $C_{1}(0)=\sqrt{\frac{4}{7}}$, $C_{2}(0)=\sqrt{\frac{3}{7}}$, while in (c)
$C_{1}(0)=\sqrt{\frac{4}{7}}$, $C_{2}(0)=-i\sqrt{\frac{3}{7}}$.
In all the (a), (b) and (c), the parameters characterizing the
initially uncorrelated state of atom-mode are
$B_{1}(0)=C_{2}(0)$, $B_{2}(0)=C_{1}(0)$. The top
horizontal line in the figures marks the upper bound due to Eq. (\ref{bound}).
}\label{fig:2}
\end{figure}

In Fig.\ref{fig:2}~(a), (b) and (c),  we plot the dynamics of the trace distance  in Eq.~(\ref{distance})
without dissipation, i.e., for $\Gamma=0$ and with dissipation for various values of $\Gamma/\Omega$,
and we consider, respectively, $\theta=\frac{\pi}{2}, 0$ and $\frac{3\pi}{2}$ for the correlated initial atom-mode state (\ref{corr}).
For the uncorrelated state (\ref{prod}), we set $B_{1}(0)=C_{2}(0)$, $B_{2}(0)=C_{1}(0)$, the same as in \cite{JCM}.
The trace distance of atomic states
can exceed its initial value for both $\Gamma=0$ and
$\Gamma/\Omega\neq0$, however only in the former case it can
reach the upper bound in Eq.~(\ref{bound}) in the course of time, see Fig.\ref{fig:2}~(a) and (c). 
{In the presence of dissipation,} because of the coupling with a larger environment as in Eq.~(\ref{mast}), the information flowing from the atom to the mode
can flow into the remaining part of the structured reservoir, as well. But then, it can no longer flow
back to the atom-mode system since the mode leaks into a Markovian reservoir, i.e., there is a unidirectional
flow of information from the mode to the reservoir \cite{blp}.
The distinct behaviors of the trace distance in weak and strong coupling regimes are exhibited, respectively,
by asymptotical decay and damped oscillations.

{In Fig.\ref{fig:2}, one can observe the
crucial influence of the initial relative phase on the dynamics of the trace distance between atomic states and, in particular, 
on the maximum value attained, i.e.,
the maximum amount of information that is accessible through measurements performed
on the open system only \cite{JCM}. This is the case both for the dissipative situation and for $\Gamma=0$,
where the presence of an initial relative phase allows to reach the upper bound in Eq.(\ref{bound}).}
In the case of a vanishing relative phase, that is shown in Fig.\ref{fig:2} (b),
the maximum value of the trace distance between atomic states
is in fact substantially smaller than the upper bound in
Eq.~(\ref{bound}), even for $\Gamma=0$.
{Note that, unlike the excitation transfer described in the previous section,
the back flow of information to the atom, which is reflected into the maximum value
reached by the trace distance, is strongly amplified by both constructive and 
destructive interference.
Indeed, this traces back to the specific dependence of the information flow
between the atom and the mode on the  the atomic excited-state populations
with and without
initial correlations, see Eq.~(\ref{distance}).
In addition, one can easily find several examples showing how the initial
relative phase plays an indispensable role also in determining whether the trace
distance of atomic states can actually increase above its initial value, which is indeed
a priori not guaranteed by the condition $I[\rho_{AM}^{(1)}(0),\rho_{AM}^{(2)}(0)]>0$.}

\section{Enhancement of entanglement}\label{sec:eoe}
In order to give a further evidence of the role of the initial relative phase,
we focus now on the dynamics of the entanglement between two atoms interacting with a {}{common} structured
reservoir.
The effect of initial correlations on the entanglement dynamics have been studied in \cite{en1, en2}.

Consider two identical two-level atoms $A$ and $B$ interacting with a {}{common}
mode of the radiation field, that is in turn coupled to a damping reservoir.
We further assume that the atoms-mode dynamics
is determined by the Lindblad equation (\ref{mast}),
where $\rho$ now represents the density operator of the system formed by the two atoms and the
mode.
The Hamiltonian $
{H}={H}_{0}+{H}_{\mathrm{int}}+{H}_{dd}$ is given by the sum of
the free Hamiltonian,
$ {H}_{0}=\omega _{0}{\sigma}_{+}^{A}{\sigma}_{-}^{A}+\omega _{0}{\sigma}_{+}^{B}{\sigma}_{-}^{B}
+\omega _{c}{b}^{\dag}{b}$,
the coupling term between the atoms and the mode
$
{H}_{\mathrm{int}}=\Omega[({\sigma}_{-}^{A}+{\sigma}_{-}^{B}){b}^{\dag}+
({\sigma}_{+}^{A}+{\sigma}_{+}^{B}){b}],
$
and the term describing dipole-dipole interaction of the atoms,
$
{H}_{dd}=D({\sigma}_{+}^{(A)}{\sigma}_{-}^{(B)}+
{\sigma}_{-}^{(A)}{\sigma}_{+}^{(B)}).
$
Indeed, ${\sigma}_{\pm}^{j}$ ($j=A,B$) are the raising and lowering operators, $\omega _{0}$ is the transition
frequency of the atom, $\Omega$
is the coupling constant between the atoms and the mode and $D$ is the coupling strength of the
two atoms.
Here, we take into account a correlated initial state
$\rho_{ABM}(0)=\left|\Psi(0)\right\rangle_{ABM}\left\langle\Psi(0)\right|$,
where
\begin{equation}\label{corr3}
\left|\Psi(0)\right\rangle_{ABM}=C_{1}(0)\left|e,g,0\right\rangle_{ABM}+C_{2}(0)\left|g,e,0\right\rangle_{ABM}+
C_{3}(0)\left|g,g,1\right\rangle_{AM}.
\end{equation}
Since there is only one
excitation in the total system, we proceed as in Sec.\ref{sec:ms} and make
the ansatz that the atoms-mode state at time $t$ is of the form
$
\rho_{ABM}(t) = |\tilde{\psi}(t)\rangle_{ABM} \langle \tilde{\psi}(t)| + \lambda(t) |g,g,0\rangle_{ABM}\langle g,g,0|,
$
with $0\leq \lambda(t) \leq 1$, $\lambda(0)=0$ and
\begin{equation}\label{eq:unno2}
|\widetilde{\psi}(t)\rangle_{ABM} =  \widetilde{C}_1(t) |e,g,0\rangle_{AM} +
\widetilde{C}_2(t) |g,e,0\rangle_{ABM}+\widetilde{C}_3(t) |g,g, 1 \rangle_{ABM}.
\end{equation}
The dynamics of the unnormalized atoms-mode
state in Eq.~(\ref{eq:unno2}) is fixed by:
\begin{eqnarray}
i\frac{d}{dt}\widetilde{C}_{1}(t)&=&\omega \widetilde{C}_{1}(t)+\Omega \widetilde{C}_{3}(t)+D\widetilde{C}_{2}(t),\nonumber\\
i\frac{d}{dt}\widetilde{C}_{2}(t)&=&\omega \widetilde{C}_{2}(t)+\Omega \widetilde{C}_{3}(t)+D\widetilde{C}_{1}(t),\nonumber\\
i\frac{d}{dt}\widetilde{C}_{3}(t)&=&\left(\omega- i \frac{\Gamma}{2}\right)\widetilde{C}_{3}(t)+\Omega \widetilde{C}_{1}(t)
+\Omega \widetilde{C}_{2}(t),\label{eq:garra}
\end{eqnarray}
while $\lambda(t)$ satisfies
$
d \lambda(t) / d t = \Gamma | \widetilde{C}_3(t) |^2.
$
To quantify the entanglement of the two atoms, we adopt Wootters' concurrence \cite{con},
which for any two-qubits density matrix $\rho$ is defined as
$
C(\rho )=\max \{0,\sqrt{\lambda _{1}}-\sqrt{\lambda _{2}}-\sqrt{\lambda _{3}}%
-\sqrt{\lambda _{4}}\}, $
where $\lambda _{i}$ ($\lambda _{1}\geq \lambda _{2}\geq \lambda _{3}\geq
\lambda _{4})$ are the eigenvalues of the matrix $\zeta =\rho (\sigma
_{y}\otimes \sigma _{y})\rho ^{*}(\sigma _{y}\otimes \sigma _{y}),$ with $%
\sigma _{y}$ the Pauli matrix and $\rho ^{*}$ the complex conjugation
of $\rho $ in the standard basis. From Eq (\ref{eq:unno2}),
one has for the concurrence of atoms $A$ and $B$
\begin{equation}\label{CAB}
C_{AB}(t)=2|\widetilde{C}_1(t)\widetilde{C}_2(t)|.
\end{equation}
In the following, we use the notation
$ C_{1}(0)=\mathcal{C}_{1}(0), C_{2}(0)=\mathcal{C}_{2}(0)e^{i\theta_{1}},
 C_{3}(0)=\mathcal{C}_{3}(0)e^{i\theta_{2}},
$
where $\mathcal{C}_{1}(0)$, $\mathcal{C}_{2}(0)$ and $\mathcal{C}_{3}(0)$ are real numbers and
$\theta_{1},\theta_{2}\in[0,2\pi]$.

\begin{figure}
\centering
\begin{minipage}[b]{2.5in}
\centering
\includegraphics[width=3in]{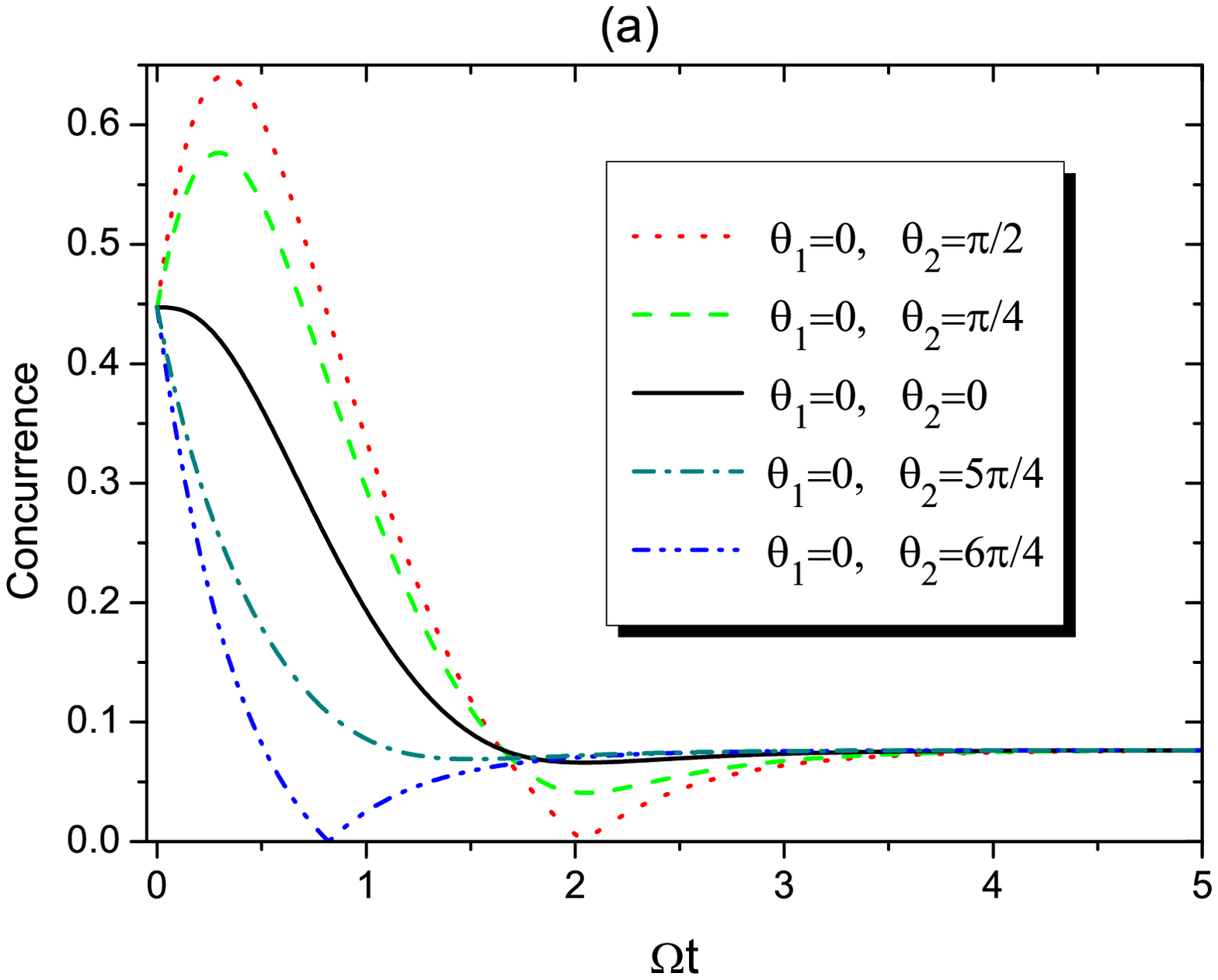}
\end{minipage}%
\begin{minipage}[b]{2.5in}
\centering
\includegraphics[width=3in]{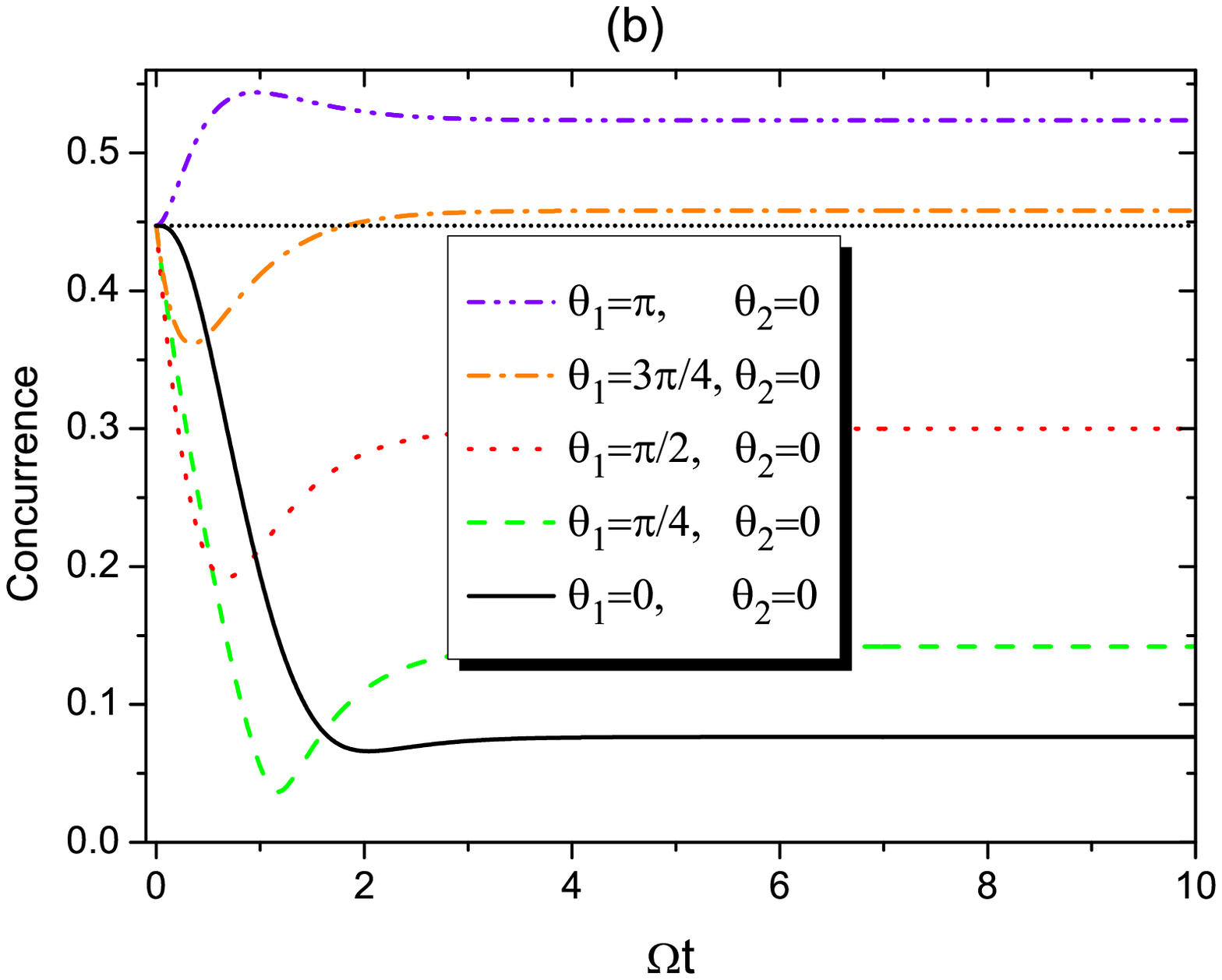}
\end{minipage}
\caption{The evolution of the concurrence $C_{AB}(t)$ of atoms $A$ and $B$
for the correlated initial state $\rho_{ABM}(0)$ with various
relative phases $\theta_{1}$ and $\theta_{2}$. The solid line is referred to the case with $\theta_1 = \theta_2 = 0$.
The parameters characterizing the initial correlated state
are $\mathcal{C}_{1}(0)=\sqrt{\frac{1}{2}}$,
$\mathcal{C}_{2}(0)=\sqrt{\frac{1}{10}}$ and
$\mathcal{C}_{3}(0)=\sqrt{\frac{4}{10}}$. The other parameters are $\Gamma/\Omega=6$
and $D=0$. }\label{fig:3}
\end{figure}

In Fig.\ref{fig:3} (a) and (b), we plot the time-evolution
of the concurrence $C_{AB}(t)$
for the correlated initial state (\ref{corr3}), with various
relative phases $\theta_{1}$ and $\theta_{2}$. We set $D=0$ in order
to focus on the entanglement dynamics due to the interaction with the environment.
Fig.\ref{fig:3} (a) shows the variations of $C_{AB}(t)$ for different $\theta_{2}$
with a fixed $\theta_{1}=0$.
It is worth noting that the entanglement can exhibit a temporary growth over its initial
value, e.g. for $\theta_{2}=\pi/4, \pi/2$, and the amplitude of the growth is determined by $\theta_{2}$.
We also note that, though the phase $\theta_{2}$ affects
the dynamics of atomic entanglement, the steady entanglement
does not depend on it. On the other hand, as shown in Fig.\ref{fig:3} (b),
steady entanglement can be increased by adjusting the relative
phase $\theta_{1}$ with a fixed $\theta_{2}$ and, in particular, it grows through changing the initial phase
$\theta_1$ from $0$ to $\pi$. Interestingly, the steady entanglement can
be greater than the initial entanglement for some values of $\theta_{1}$ (e.g., here $\theta_{1}=3\pi/4, \pi$). Moreover, for $\theta_{1}=\pi$ and $\theta_{2}=0$,
atomic entanglement starts increasing at the initial time and then it takes values
that are always higher than the initial one. The physical reason for these results can still be attributed to
the quantum interference induced by the initial relative phases. The only difference between the present situation
and that of a single atom is the increased amount of processes involved in
quantum interference.

\section{Conclusion}\label{sec:c}
We have investigated the role of the relative phase of
a correlated initial total state in the subsequent open-system dynamics.
In particular, we have taken into account the dynamics of a two-level atom interacting with a
structured reservoir. The initial relative phase affects
the two processes of excitation transfer, respectively, from the atom to the mode and
from the mode to the atom.
We have further shown how this reflects into the
dynamics of the information flow between the atom and the reservoir.
The quantum interference induced by an initial relative phase, in fact, strongly influences the dynamics of the trace distance of atomic states
and in particular the maximum value that it actually
reaches during the dynamics.
Finally, we have considered two two-level atoms interacting
with the same structured reservoir. We have shown how relative phases in the initial correlated total state
can enhance atomic entanglement.
As a final remark, let us note that the sensitivity of open system dynamics on
the initial relative phase suggests possible
ways to detect it through measurements on the open system only \cite{trace2}.
The trace-distance analysis of reduced dynamics allows
to access, apart from overall correlation properties \cite{blp, gb},
specific features of the initial total state, which can be useful,
e.g., when this is only partially controlled during the preparation
procedure.\\

{\textbf{Acknowledgments}}

We thank Prof. Masashi Ban for useful comments
and suggestions. This work is supported by National Natural
Science Foundation of China under Grant Nos. 10947006 and 61178012,
the Specialized Research Fund for the Doctoral Program of Higher
Education under Grant No. 20093705110001,
Scientific Research Foundation of Qufu Normal University for Doctors, MIUR
under PRIN 2008, and COST under MP1006.

\clearpage


\begin{thebibliography}{99}
\bibitem{open} H.-P. Breuer, F. Petruccione, The Theory of Open Quantum
Systems, Oxford University Press, Oxford, 2002.
\bibitem{qci} M.A. Nielsen, I.L. Chuang, Quantum Computation and Quantum Information,
Cambridge University Press, Cambridge, 2000.

\bibitem{ini1} P. Pechukas, Phys. Rev. Lett. {73} (1994)  1060.
\bibitem{ini2} L. D. Romero, J.P. Paz, Phys. Rev. A {55} (1997) 4070.
\bibitem{ini4} P. \v{S}telmachovi\v{c}, V. Bu\v{z}ek, Phys. Rev. A {64} (2001) 062106.
\bibitem{ini5} T.F. Jordan, A. Shaji, E.C.G. Sudarshan, Phys. Rev. A {70} (2004) 052110.
\bibitem{ini7} M. Ban, Phys. Rev. A {80} (2009) 064103.
\bibitem{ini10} Y. J. Zhang, X. B. Zou, Y. J. Xia, G. C. Guo, Phys. Rev. A
{82} (2010) 022108.
\bibitem{ini11} A. G. Dijkstra, Y. Tanimura, Phys. Rev. Lett. {104} (2010) 250401.
\bibitem{ini12} H. T. Tan, W. M. Zhang, Phys. Rev. A {83} (2011) 032102.
\bibitem{ini13} B. Arend G. Dijkstra, Y, Tanimura, arXiv:1111.3722v1 [quant-ph].
\bibitem{ini14} D. Z. Rossatto, T. Werlang, L. K. Castelano, C. J. Villas-Boas, F. F. Fanchini,
Phys. Rev. A {84} (2011) 042113.
\bibitem{ini15} M. Ban, S. Kitajima, F. Shibata, Int. J. Theor. Phys.
DOI: 10.1007/s10773-012-1121-y.


\bibitem{trace1} E.-M. Laine, J. Piilo, H.-P. Breuer, Eur. Phys. Lett. {92} (2010) 60010.
\bibitem{trace3} J. Dajka, J. Luczka, Phys. Rev. A {82} (2010) 012341;
J. Dajka, J. Luczka, P. H\"{a}nggi, Phys. Rev. A {84} (2011) 032120.
\bibitem{JCM} A. Smirne, H. P. Breuer, J. Piilo, B. Vacchini, Phys. Rev. A
{82} (2010) 062114.
\bibitem{trace2} C. F. Li, J. S. Tang, Y. L. Li, G. C. Guo Phys. Rev. A {83} (2011) 064102; \,
A. Smirne, D. Brivio, S. Cialdi, B. Vacchini, M.G.A. Paris, Phys. Rev. A {84} (2011) 032112.

\bibitem{cohen} C. Cohen-Tannoudji et al., Atom-Photon Interactions, John Wiley, New York, 1998.

\bibitem{damp} M. Scala, B. Militello, A. Messina, S. Maniscalco, J. Piilo, K.-A. Suominen, J. Phys. A {40} (2007) 14527.

\bibitem{pseu1} B. M. Garraway, Phys. Rev. A {55} (1997) 2290.

\bibitem{pseudo} L. Mazzola, S. Maniscalco, J. Piilo, K. A. Suominen,
B.M.Garraway, Phys. Rev. A {79} (2009) 042302.




\bibitem{blp}H.-P.~Breuer, E.-M.~Laine, J.~Piilo,
   Phys. Rev. Lett. {103} (2009) 210401; \, E.-M.~Laine, J.~Piilo, H.-P.~Breuer,
  Phys. Rev.~A  {81} (2010) 062115.



\bibitem{en1} M. Ban, S. Kitajima, F. Shibata, Phys. Lett. A {375} (2011) 2283.
\bibitem{en2} L. Li, J. Zou, Z. He, J.-G. Li, B. Shao, L.-A. Wu, Phys. Lett. A {376} (2012) 913.

\bibitem{con} W. K. Wootters, Phys. Rev. Lett. {80} (1998) 2245.

\bibitem{gb} M. Gessner, H.-P. Breuer, Phys. Rev. Lett. {107} (2011) 180402.


\end{thebibliography}
\end{document}